\documentclass{IEEEtran4PSCC}
\usepackage{cite}
\ifCLASSINFOpdf
   \usepackage[pdftex]{graphicx}
\else
   \usepackage[dvips]{graphicx}
\fi
\usepackage{amsthm}
\usepackage[framemethod=tikz]{mdframed}
\usepackage[cmex10]{amsmath}
\usepackage{amsmath,bm}
\usepackage{bbm}
\usepackage{url}
\usepackage{tabularx}
\usepackage{colortbl,booktabs}
\usepackage{multirow}
\usepackage{threeparttable}
\usepackage{makecell}
\usepackage{algorithm}
\usepackage{verbatim}
\usepackage{siunitx}
\usepackage{xcolor}
\usepackage{diagbox}
\usepackage{enumerate}
\usepackage{threeparttable}
\usepackage{graphicx}
\usepackage{subfigure}
\usepackage{algpseudocode}
\usepackage{makecell}
\usepackage{nicematrix}
\usepackage{amsfonts}
\usepackage{hyperref}

\newtheorem{assumption}{\text{\textbf{Assumption}}}
\newtheorem{theorem}{Theorem}

\newtheorem{proposition}{Proposition}
\newtheorem{definition}{Definition}

\makeatletter
\let\old@ps@headings\ps@headings
\let\old@ps@IEEEtitlepagestyle\ps@IEEEtitlepagestyle
\def\psccfooter#1{%
    \def\ps@headings{%
        \old@ps@headings%
        \def\@oddfoot{\strut\hfill#1\hfill\strut}%
        \def\@evenfoot{\strut\hfill#1\hfill\strut}%
    }%
    \def\ps@IEEEtitlepagestyle{%
        \old@ps@IEEEtitlepagestyle%
        \def\@oddfoot{\strut\hfill#1\hfill\strut}%
        \def\@evenfoot{\strut\hfill#1\hfill\strut}%
    }%
    \ps@headings%
}
\makeatother

\psccfooter{%
        \parbox{\textwidth}{\hrulefill \\ \small{23nd Power Systems Computation Conference} \hfill \begin{minipage}{0.2\textwidth}\centering \vspace*{4pt} \includegraphics[scale=0.06]{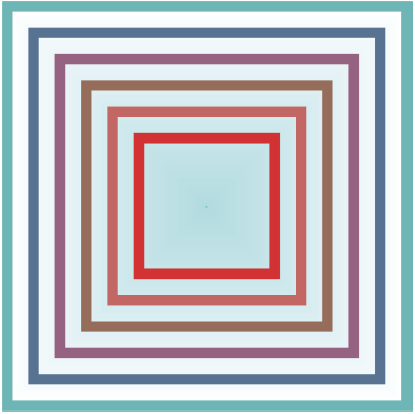}\\\small{PSCC 2024} \end{minipage} \hfill \small{Paris-Saclay, France --- June 4th -- 7th, 2024}}%
}

\begin{document}

\title{Deriving Loss Function for Value-oriented Renewable Energy Forecasting}

%% To specify the authors when (number of affiliations <= 2)
% \author{
% \IEEEauthorblockN{Author n.1 Name per Affiliation A\\ Author n.2 Name per Affiliation A}
% \IEEEauthorblockA{(Affiliation A) Department Name of Organization \\
% Name of the organization, acronyms acceptable\\
% City, Country\\
% \{email author n.1, email author n.2\}@domain (if desired)}
% \and
% \IEEEauthorblockN{Author n.1 Name per Affiliation B\\ Author n.2 Name per Affiliation B}
% \IEEEauthorblockA{(Affiliation B) Department Name of Organization \\
% Name of the organization, acronyms acceptable\\
% City, Country\\
% \{email author n.1, email author n.2\}@domain (if desired)}
% }

%% To specify the authors when (number of affiliations > 2)
% \author{\IEEEauthorblockN{Author n.1\IEEEauthorrefmark{1},
% Author n.2\IEEEauthorrefmark{2},
% Author n.3\IEEEauthorrefmark{3}, 
% Author n.4\IEEEauthorrefmark{3} and
% Author n.5\IEEEauthorrefmark{4}}
% \IEEEauthorblockA{\IEEEauthorrefmark{1} Department Name of Organization A\\
% Name of the organization A,
% Address A\\ Emails if wanted}
% \IEEEauthorblockA{\IEEEauthorrefmark{2} Department Name of Organization B\\
% Name of the organization B,
% Address B\\ Emails if wanted}
% \IEEEauthorblockA{\IEEEauthorrefmark{3} Department Name of Organization C\\
% Name of the organization C,
% Address C\\ Emails if wanted}
% \IEEEauthorblockA{\IEEEauthorrefmark{4}Department Name of Organization D\\
% Name of the organization D,
% Address D\\ Emails if wanted}
% }

\author{\IEEEauthorblockN{
Yufan Zhang\IEEEauthorrefmark{1},
Honglin Wen\IEEEauthorrefmark{2}, Yuexin Bian\IEEEauthorrefmark{1}, and
Yuanyuan Shi\IEEEauthorrefmark{1}}
\IEEEauthorblockA{\IEEEauthorrefmark{1} Department of Electrical and Computer Engineering, University of California San Diego, San Diego, CA, USA}
\IEEEauthorblockA{\IEEEauthorrefmark{2} Department of Electrical Engineering, Shanghai Jiao Tong University, Shanghai, China}
}

\maketitle

\begin{abstract}
Renewable energy forecasting is the workhorse for efficient energy dispatch. However, forecasts with small mean squared errors (MSE) may not necessarily lead to low operation costs. Here, we propose a forecasting approach specifically tailored for operational purposes, by incorporating operational problems into the estimation of forecast models via designing a \emph{loss function}. We formulate a bilevel program, where the operation problem is at the lower level, and the forecast model estimation is at the upper level. We establish the relationship between the lower-level optimal solutions and forecasts through multiparametric programming. By integrating it into the upper-level objective for minimizing expected operation cost, we convert the bilevel problem to a single-level one and derive the loss function for training the model. It is proved to be piecewise linear, for linear operation problem. Compared to the commonly used loss functions, e.g. MSE, our approach achieves lower operation costs.

\end{abstract}


\section{Introduction}
Forecasting is usually regarded as an indispensable tool to accommodate the uncertainty of renewable energy sources (RESs). It leverages information at the current time and predicts the generation at future time. Typically, the forecast model can be developed using data-driven methods, especially cutting-edge techniques such as deep learning \cite{xiong2022short} and gradient boosting machines \cite{landry2016probabilistic}.
At the model estimation (training) stage, a loss function is required to guide the optimization of the forecast model parameters. For instance, mean squared error (MSE) and pinball loss \cite{gneiting2011quantiles} are respectively used as loss functions for point and quantile forecasting. At the operation stage, forecasts are issued via the trained model and fed into the subsequent operation problem as inputs, which is referred to as the ``predict, then optimze'' pipeline. 
Till now, several models have been developed to improve the statistical quality of forecasts; see a comprehensive review \cite{hong2020energy}.

Unsurprisingly, forecasts can impact decisions, and therefore impact the value of the operation problem. %Although the definition of value may differ case-by-case, 
Previous studies show that there is no guarantee that forecasts with good statistical quality lead to higher value in operations \cite{10185994,stratigakos2022prescriptive,carriere2019integrated} (e.g., no guarantee for lower operational costs). It has been suggested in \cite{christoffersen1997optimal} that strategic forecasts are preferred when the penalty on forecast deviation is asymmetric. Therefore, it is promising to design forecast models with high operational value instead of high accuracy. Here, we refer to such forecasts as \textit{value-oriented forecasts} and the traditional ones as \textit{quality-oriented forecasts}. 

Typically, the most salient challenge to developing value-oriented forecast models lies in model training.
For that, one idea is to integrate the forecast model into the downstream operation problem and estimate the parameters by solving the integrated program, with the assumption that forecast model is linear \cite{chen2021feature}. Another idea is to encode the operation problem as a differentiable optimization layer~\cite{donti2017task,wahdany2023more}, and to estimate the forecast model together with solving the operation problem in an end-to-end fashion. 
However, they all confine forecast models to specific types \cite{chen2021feature,donti2017task,wahdany2023more}, the former requires the forecast models to be linear while the latter requires models to be differentiable, which hinders the applicability. 

A more general idea is to design loss functions for forecast model estimation that capture the value of the forecast in subsequent operational problems. 
%Then, forecast models can be flexibly set as linear or nonlinear models such as neural networks and still trained by minimizing the loss function. 
For instance, the regret loss has been widely used \cite{mandi2023decision}, which is defined as the difference between the optimal objective values under the realization and the forecast. With this idea, the ``Smart Predict, then Optimize" (SPO) loss is proposed in~\cite{elmachtoub2022smart}, with a special focus on single-stage operation problems with linear objectives where the cost vector contains unknown parameters \cite{stratigakos2021value}.
However, it may have limited applicability in the context of sequential power systems operation, where the value of forecast is related with multiple correlated operation problems. Concretely, most power systems operation problems are organized into two stages \cite{kirschen2018fundamentals}: at the day-ahead stage, the generation schedule of slow-start generators based on RES forecasts is established; while at the real-time stage close to the actual delivery, any imbalance from the day-ahead schedule is settled by scheduling flexible resources.
%Some studies \cite{zhao2022uncertainty,morales2014electricity} show that the quality-oriented forecasts, typically conditional expectations, result in a lack of coordination between the two stages, and therefore lead to high operation cost. 
Therefore, it remains an open issue to design value-oriented forecasting approach for use in such sequential operation problems.

In contrast to the approach taken in \cite{li2016toward,zhang2022cost} in which they design a heuristic~\cite{li2016toward} or empirical~\cite{zhang2022cost} loss function, we propose to \emph{theoretically} derive a loss function for value-oriented forecasting. The special focus is placed on the virtual power plant (VPP) operator with wind power, which performs the day-ahead and real-time energy dispatch. The wind power forecast is issued for the day-ahead problem, and the real-time problem is performed to settle the forecast deviation. In this line, the value of the forecast is related to the sum of operation costs of the day-ahead and real-time problems. Similar to our previous work \cite{zhang2023valueoriented}, we formulate the task of forecast model parameter estimation as a bilevel program, where the lower level solves the day-ahead and real-time problems given the forecasts issued by the upper level. Based on the optimal solutions provided by the lower level, the forecast model parameters at the upper level are estimated, for minimizing the expected total operation costs at two stages.
% \todo{It's good that we first mentioned similar to our previous work [18]... but what is missing here is the difference compared to [18]. We need to explicitly say somethings like ``different than [18] (summarize how [18] did to solve the bi-level optimization), this paper... (summarize how our paper solve the bi-level optimization)" and justify the benefits of this new method}

% \todo{I am not sure whether it would be better to have a more distinguishable part for summarizing the contributions (like itemized) than the current version. Maybe have a try for both versions and decide}

Different from \cite{zhang2023valueoriented} where only a local loss function in the neighborhood of a given sample is obtained, in this work, we derive the analytical form of loss function across the entire domain of forecast values via multiparametric programming~\cite{Gal+2010}. Concretely, under the assumption that the day-ahead and real-time energy dispatch problems are linear programs (LPs), the relationship between the forecasts and the optimal solutions to lower level problems is piecewise linear. By plugging it into the upper level objective, the loss function is obtained and the bilevel program is transformed to a single-level one. We show such a loss function is piecewise linear. With the derived loss function, we no longer need to iteratively solve the operation problems for different samples at the training stage, thus significantly improving the computational efficiency. We note that analytically characterizing the piecewise linear loss function for value-oriented forecasting is the main contribution of this work.
%The main contribution is summarized as follows,

% \textcolor{blue}{We note that analytically characterizing the piecewise linear loss function for value-oriented forecasting is the main contribution of this work.}
%In general, the optimal forecast under such a loss function cannot be analytically derived \cite{christoffersen1997optimal}, except for its special case, i.e., the pinball loss. Also, the usually used forecasts, i.e., conditional expectations, are always the sub-optimal \cite{christoffersen1997optimal}.

% The main contribution of the paper is theoretically design a loss function for value-oriented forecasting, which is proved to be piecewise linear for the operation problem in the form of LP.

%1) A single-level reduction to solve the bilevel program for estimating the model parameters, where the relationship between the lower level optimal operation solutions and the forecast is explicitly derived.

% \textcolor{blue}{A loss function for value-oriented forecasting is theoretically derived, which is proved to be a piecewise linear function for the operation problem in the form of LP.}

% \textcolor{blue}{2) A single-level reduction to solve the bilevel program for estimating the model parameters.}

The remaining parts of this paper are organized as follows. The preliminaries regarding the day-ahead and real-time energy dispatch problems, and the bilevel program at the training stage are given in Section \uppercase\expandafter{\romannumeral2}. Section \uppercase\expandafter{\romannumeral3} derives the loss function for value-oriented forecasting and presents the solution strategy. Results are discussed and evaluated in Section \uppercase\expandafter{\romannumeral4}, followed by the conclusions.

%\vspace{0.1cm}
\textit{Notations:} Variables are denoted as letters in lowercase, e.g. $x$, while vectors are denoted in bold, e.g. $\boldsymbol{x}$. The all-one and all-zero vectors are denoted as $\bm{1}$ and $\bm{0}$. $\bm{O}$ is an all-zero matrix with a size to be defined by the use case. In particular, the variables in day-ahead problem are indexed by the subscript $D$, whereas the variables in the real-time problem are indexed by the subscript $R$.

\section{Preliminaries}
We introduce the energy dispatch model for VPP operators with wind power, solved at the day-ahead and real-time stages in subsection \textit{A}. Then, the parameter estimation of the forecast model, for minimizing the expected day-ahead and real-time operation costs is presented in subsection \textit{B}.

\begin{figure}[h]
  \centering
  \includegraphics[scale=0.6]{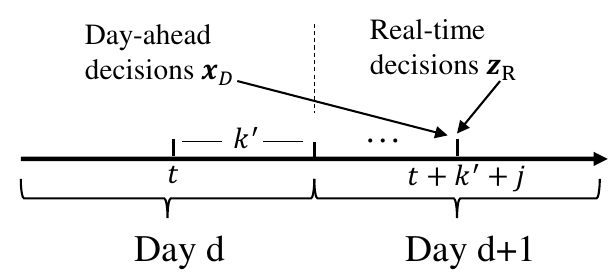}\\
  \caption{The timeline of the day-ahead and real-time energy dispatch problems.}\label{Fig 1}
\end{figure}

\subsection{Day-ahead and Real-time Energy Dispatch Models}

The VPP operator manages a non-networked system, in charge of wind power, slow-start generators (SGs), and flexible resources. The energy dispatch of the day-ahead and real-time problems is performed in a sequential order. The timeline of the two problems is shown in Fig. \ref{Fig 1}. The day-ahead problem is solved at time $t$ on day $d$, and determines the power generation of the SGs to be delivered at time $t+k^\prime+j,\forall j=0,...,23$ on day $d+1$, where $k^\prime$ is the time interval between $t$ and \textit{0 a.m.} on day $d+1$. Wind power forecasts are issued at $t$ on day $d$ as well, for the future time $t+k^\prime+j,\forall j=0,...,23$ on day $d+1$. Given the contextual information $\bm{s}_{t+k^\prime+j}$ for the time-slot $t+k^\prime+j$, the wind power forecast $\hat{y}_{t+k^\prime+j}$ is issued by a forecast model $g$ with parameters $\Theta$,
\begin{equation}\label{1}
    \hat{y}_{t+k^\prime+j}=g(\bm{s}_{t+k^\prime+j};\Theta),\forall j=0,...,23.
\end{equation}

Throughout this paper, we focus on operation problems that can be modeled by linear programs. The ramping constraints are not considered, and therefore the day-ahead problem can be solved for each time-slot independently. We use the instance solved for time-slot $t+k^\prime+j$ as an example, and drop the time index in each variable for notational simplicity.  Let $\bm{x}_D$ denote the scheduling decisions of SGs. The day-ahead problem solved at time $t$, with the linear generation cost $\bm{c}_D^\top\bm{x}_D$, is,
\begin{subequations}\label{2}
\begin{alignat}{2}
&\mathop{\min}_{\bm{x}_{D}} 
&& \quad \bm{c}_D^\top  \bm{x}_{D}\label{2a}\\
    &\text{s.t.} &&  \quad  \bm{A}_D\bm{x}_{D}\leq \bm{b}_D\label{2b}\\
    &&& \quad \bm{1}^\top  \bm{x}_{D}+\hat{y}=l,\label{2c}
\end{alignat}
\end{subequations}
where $\bm{c}_D,\bm{A}_D,\bm{b}_D$ are the known coefficients. \eqref{2b} includes the upper and lower bounds of the generation power of SGs. Given the load $l$, whose forecast is assumed to be rather accurate and incurs no uncertainty, \eqref{2c} enforces the power balance between the supply (dispatch from SGs and the wind power production) and the load. 

% \textcolor{blue}{$\lambda$ is the dual variable associated with the constraint \eqref{2c}, whose optimal solution $\lambda^*$ can be interpreted as the price of generation for supplying the demand and equal to the marginal generation cost of the marginal SG.}

Once the wind power realization $y$ is revealed at time $t+k^\prime+j$ on day $d+1$, the real-time problem is solved at that time-slot to deal with the energy imbalance $y-\hat{y}$, by leveraging the flexible resources whose outputs are denoted as $\bm{z}_R$. Minimizing the linear cost $\bm{c}_R^\top \bm{z}_{R}$ of the flexible resources outputs, the real-time problem is, 
\begin{subequations}\label{3}
\begin{alignat}{2}
 &\mathop{\min}_{\bm{z}_{R}} && \quad \bm{c}_R^\top  \bm{z}_{R}\label{3a}\\
    &\text{s.t.} &&  \quad  \bm{A}_R\bm{z}_{R}\leq \bm{b}_R\label{3b}\\
   &&& \quad \bm{d}_R^\top  \bm{z}_{R}+y-\hat{y}=0,\label{3c}
\end{alignat}
\end{subequations}
where $\bm{c}_R,\bm{A}_R,\bm{b}_R,\bm{d}_R$ are the coefficients. \eqref{3b} limits the flexible resources outputs within the upper and lower bounds, and \eqref{3c} ensures that the system remains in balance. 

% \textcolor{blue}{$\nu$ is the dual variable associated with the constraint \eqref{3c}, whose optimal solution $\nu^*$ can be interpreted as the price of using flexible resources to settle down the deviation and equal to the marginal cost of the marginal flexible resources.}

Here, we make the following assumption. 

% \todo{In Assumption 1, the definition for the day-ahead marginal generation cost and real-time marginal generation cost are not defined. Maybe define the dual variable for (2c) and (3c) first}

% \begin{assumption}\label{assumption 1}
% \textcolor{blue}{%The real-time marginal generation cost is different, depending on the imbalance sign. 
% There is $\lambda^* > \nu^*$, when the imbalance $y-\hat{y} >0 $; otherwise $\lambda^* < \nu^*$. }
% \end{assumption}
% This assumption ensures that the profit of the wind power $\lambda^*\hat{y}+\nu^*(y-\hat{y})$ achieves the maximum when the forecast $\hat{y}$ perfectly matches the realization $y$. As shown in \cite{zhang2023valueoriented} that maximizing wind power profit is compatible with minimizing the sum of the system operation costs of \eqref{2} and \eqref{3}, any deviation $y-\hat{y}$ will bring an increase in the operation cost.

% \todo{We need to explain what this assumption requires in a more intuitive way and justify why this is a reasonable assumption right after assumption}

\begin{assumption}\label{assumption 2}
The linear programs in \eqref{2} and \eqref{3} are neither primal nor dual degenerate. In other words, the optimal primal and dual solutions for \eqref{2} and \eqref{3} are unique.  %\todo{This assumption is already stated in Theorem 1, do we still need a seperate assumption here?}
\end{assumption}

\subsection{Training and Operation Stages}

\subsubsection{Training Stage}
At the training stage, the parameters $\Theta$ of the model $g$ are estimated. Given the training set $\{\bm{s}_m,y_m\}_{m=1}^M$ regarding the contextual information and the realization, a forecast $\hat{y}_m=g(\bm{s}_m;\Theta)$ is issued for each sample. The commonly used point forecasting approach obtains the estimated forecast model parameters $\hat{\Theta}$ by minimizing a quality-oriented loss, e.g., the MSE, at the training stage, 
\begin{subequations}\label{mse}
    \begin{align}       \tag{Quality-oriented}
    &\mathop{\min}_{\Theta} \frac{1}{M}\sum_{m=1}^M(\hat{y}_m-y_m)^2\\
        &\hat{y}_m=g(\bm{s}_m;\Theta),
    \end{align}
\end{subequations}
which offers $\hat{y}$ at the expected value of the forecast. 

Previous study \cite{morales2014electricity} showed that such an approach leads to imperfect coordination between the day-ahead and real-time operation problems in \eqref{2} and \eqref{3}, therefore resulting in large system operation costs. For that, \cite{zhang2023valueoriented} formulates a bilevel program to estimate the model parameters, minimizing the expected operation cost of the day-ahead and real-time problems. %Given the training set $\{\bm{s}_m,y_m\}_{m=1}^M$ regarding the contextual information and the realization, the forecast model parameters $\Theta$ are estimated at the upper level, minimizing the expected operation cost. 
Given $\hat{y}_m$ announced by the forecast model from the upper level, the day-ahead SGs generation $\bm{x}_{m,D}^*(\hat{y}_m,l_m)$ and the real-time flexible resources output $\bm{z}_{m,R}^*(\hat{y}_m,y_m)$ at the lower level will be determined for each sample. In turn, the lower level optimal solutions $\bm{x}_{m,D}^*(\hat{y}_m,l_m)$ and $\bm{z}_{m,R}^*(\hat{y}_m,y_m)$ are passed to the upper level and form its objective. The model estimation for value-oriented forecasting is formulated as,
\begin{subequations}\label{4}
\begin{align}
\tag{Value-oriented}
    \mathop{\min}_{\Theta} \quad & \frac{1}{M}\sum_{m=1}^M [\bm{c}_D^\top \bm{x}_{m,D}^*(\hat{y}_m,l_m)+\bm{c}_R^\top \bm{z}_{m,R}^*(\hat{y}_m,y_m)]\label{4a}\\
    \text{s.t.} \quad & \hat{y}_m=g(\bm{s}_m;\Theta),\label{4b}\\
     &  \bm{x}_{m,D}^*(\hat{y}_m,l_m)=\mathop{\arg\min}_{\bm{x}_{m,D}}  \bm{c}_D^\top \bm{x}_{m,D}\label{4c}\\
    &\qquad \qquad \qquad \quad \text{s.t.} \ \bm{A}_D  \bm{x}_{m,D}\leq \bm{b}_D\label{4d}\\
    &\qquad \qquad \qquad \quad \quad \ \bm{1}^\top  \bm{x}_{m,D}+\hat{y}_m=l_m\label{4e}\\
    &  \bm{z}_{m,R}^*(\hat{y}_m,y_m)=\mathop{\arg\min}_{\bm{z}_{m,R}} \bm{c}_R^\top  \bm{z}_{m,R}\label{4f}\\
    &\qquad \qquad \qquad \quad \text{s.t.} \
    \bm{A}_{R}  \bm{z}_{m,R}\leq \bm{b}_{R}\label{4g}\\
    &\qquad \qquad \qquad \quad \quad \ \bm{d}_{R}^\top  \bm{z}_{m,R}+y_{m}-\hat{y}_m=0,\label{4h}
\end{align}
\end{subequations} 
where \eqref{4c}-\eqref{4e} and \eqref{4f}-\eqref{4h} naturally form the lower level problem by the \textit{argmin} operation, where the forecast model output $\hat{y}_m$ acts as the parameter. We express the optimal solutions of $\bm{x}_{m,D}$ as $\bm{x}_{m,D}^*(\hat{y}_m,l_m)$, since it depends on $\hat{y}_m$ and on the load $l_m$. Similarly, the optimal solutions of $\bm{z}_{m,R}$ is expressed as $\bm{z}_{m,R}^*(\hat{y}_m,y_m)$, since it depends on $\hat{y}_m$ and on the realization $y_m$. After solving the bilevel program in \eqref{4}, the estimated parameters $\hat{\Theta}$ are obtained. 

\subsubsection{Operation Stage}
At the operation stage, the ``predict-then-optimize'' decision paradigm is applicable. Take the operation problem solved for time-slot $t+k^\prime+j$ on day $d+1$ as an example. The forecast $\hat{y}_{t+k^\prime+j}$ is obtained by $\hat{y}_{t+k^\prime+j}=g(\bm{s}_{t+k^\prime+j};\hat{\Theta})$ with the trained forecast model, and serves as an input to the day-ahead energy dispatch problem in \eqref{2}. Solving \eqref{2} obtains the dispatch decisions for the slow-start generators $\bm{x}_D^*$ with operation cost $\bm{c}_D^\top \bm{x}_D^*$. After the wind power realization $y_{t+k^\prime+j}$ is revealed, solving \eqref{3} obtains the real-time dispatch decisions for the flexible resources $\bm{z}_R^*$ and
operation cost $\bm{c}_R^\top \bm{z}_R^*$. 

\section{Derivation of the Loss Function}
In this section, we theoretically derive the loss function for value-oriented forecasting. Then, we show the relationship between the proposed approach and the differentiable optimization approach proposed in \cite{donti2017task}, where the loss function is not explicitly derived.

%\subsection{Training Stage}

Here, we propose to obtain the loss function by converting the bilevel program in \eqref{4} into a single-level one. The objective of the reduced single-level program will be used as the loss function for training the forecast model. This transformation depends on the relationship between the optimization problem parameters, and the optimal solutions, which can be derived via multiparametric programming~\cite{borrelli2003geometric}. For notational simplicity, we will drop the sample index $m$ during the derivation process. For a nondegenerate linear program where the parameters of interest are in the constraints, we can explicitly characterize the relationship between the parameters and the optimal primal solutions. Before proceeding it, we first give the definition of optimal partition and critical region.
\begin{definition}\emph{\textbf{(Optimal partition of active and in-active constraints)}}
\label{def1}
Consider the linear program \eqref{lp} which is neither primal nor dual degenerate and has the optimal solution $\bm{x}^*(\bm{\theta})$, i.e.,
\begin{subequations}\label{lp}
\begin{alignat}{2}
 \bm{x}^*(\bm{\theta})=&\mathop{\arg\min}_{\bm{x}} && \quad \bm{c}^\top  \bm{x}\\
    &\text{s.t.} &&  \quad  \bm{G}\bm{x}\leq \bm{w}+\bm{F}\bm{\theta}\label{lpb}.
\end{alignat}
\end{subequations}

Let $\mathcal{J}$ denote the set of constraint indices in \eqref{lpb}. For any subset $\mathcal{J}^\prime \in \mathcal{J}$, let $\bm{G}_{\mathcal{J}^\prime},\bm{F}_{\mathcal{J}^\prime}$ be the submatrix of $\bm{G},\bm{F}$, and $\bm{w}_{\mathcal{J}^\prime}$ be the subvector of $\bm{w}$. An optimal partition of the active and in-active constraints of the index set $\mathcal{J}$ associated with the parameters $\bm{\theta}$ is the partition $(\mathcal{J}^{\bm{\theta}},\overline{\mathcal{J}^{\bm{\theta}}})$, where 
\begin{subequations}
\begin{align}
        &\bm{G}_{\mathcal{J}^{\bm{\theta}}}\bm{x}^*(\bm{\theta})=\bm{w}_{\mathcal{J}^{\bm{\theta}}}+\bm{F}_{\mathcal{J}^{\bm{\theta}}}\bm{\theta}\label{pa}\\
        &\bm{G}_{\overline{\mathcal{J}}^{\bm{\theta}}}\bm{x}^*(\bm{\theta})<\bm{w}_{\overline{\mathcal{J}}^{\bm{\theta}}}+\bm{F}_{\overline{\mathcal{J}}^{\bm{\theta}}}\bm{\theta}\label{pb}.
    \end{align}
    \end{subequations}
\end{definition}
\begin{definition}
    \emph{\textbf{(Critical region)}} Let $\mathcal{B}$ denote a region of $\bm{\theta}$. $\forall \bm{\theta} \in \mathcal{B}$, the associated active and in-active constraints are the same. That is, the optimal partition associated with $\forall \bm{\theta} \in \mathcal{B}$ is the same. The region $\mathcal{B}$ is called as the critical region.
\end{definition}

With these definitions, the following theorem characterizes the relationship between the parameters $\bm{\theta}$ and the optimal solution $\bm{x}^*(\bm{\theta})$ in the linear program \eqref{lp}.
\begin{theorem}
\label{Theorem 1}~\cite{borrelli2003geometric}
Consider the linear program \eqref{lp}, and let polyhedron set $\Lambda$ be the domain regarding $\bm{\theta}$. There exists a set of polyhedral partition $\{\mathcal{B}^i\}_{i=1}^K$ regarding the critical regions of $\Lambda$, where $i$ is the index of the critical region and $K$ is the number of critical regions, such that,
\begin{equation*}
\begin{aligned}
&\mathcal{B}^i \cap  \mathcal{B}^j=\emptyset,\forall i \neq j\\
&\cup_{i=1}^K\mathcal{B}^i=\Lambda.
\end{aligned}
\end{equation*}
In the critical region $\mathcal{B}^i$, given the associated optimal partition $(\mathcal{J}^i,\overline{\mathcal{J}^i})$, the mapping from $\bm{\theta}$ to the optimal solution of $\bm{x}$, i.e., $\bm{x}^i(\bm{\theta}),\forall \bm{\theta} \in \mathcal{B}^i, i=1,...,K$, is affine, i.e., 
\begin{equation}\label{theorem_map}
\bm{x}^i(\bm{\theta})= \bm{G}_{\mathcal{J}^i}^{-1}(\bm{w}_{\mathcal{J}^i}+\bm{F}_{\mathcal{J}^i}\bm{\theta}) 
\end{equation}
and critical region $\mathcal{B}^i$ where \eqref{theorem_map} holds is defined by,
\begin{equation}\label{theorem_region}
    \bm{G}_{\overline{\mathcal{J}}^i}\bm{G}_{\mathcal{J}^i}^{-1}(\bm{w}_{\mathcal{J}^i}+\bm{F}_{\mathcal{J}^i}\bm{\theta}) <\bm{w}_{\overline{\mathcal{J}}^i}+\bm{F}_{\overline{\mathcal{J}}^i}\bm{\theta}
\end{equation}
That is, the function $\bm{x}(\bm{\theta})=\{\bm{x}^i(\bm{\theta})\}_{i=1}^K$ is piecewise linear over $\Lambda$.
\end{theorem} 

\begin{figure}[h]
  \centering
  \includegraphics[scale=0.42]{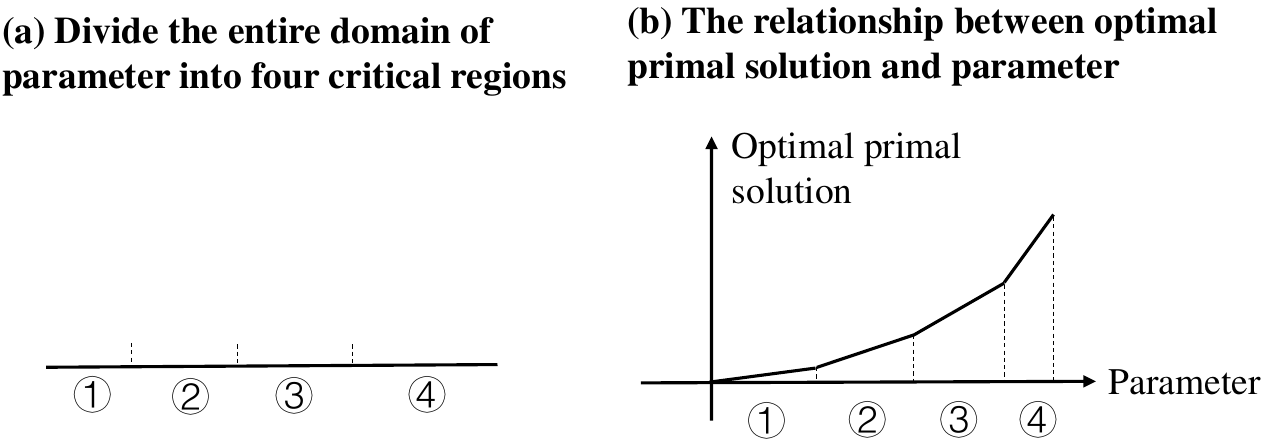}\\
  \caption{An illustration of the relationship between optimal solution of a linear program and its parameter.}\label{illustrationofmpp}
\end{figure}

Detailed proof of this classical result can be found in~\cite{borrelli2003geometric}. Here, we give a sketch of the proof. 
\begin{proof}
$\forall \bm{\theta} \in \mathcal{B}^i$, the optimal partition is the same, i.e., $\mathcal{J}^{\bm{\theta}}= \mathcal{J}^i,\overline{\mathcal{J}^{\theta}} = \overline{\mathcal{J}}^i$. Therefore, $\forall \bm{\theta} \in \mathcal{B}^i$, the associated active constraints satisfy \eqref{pa}, and the associated in-active constraints satisfy \eqref{pb}. Since  \eqref{lp} is neither primal nor dual degenerate, $\bm{G}_{\mathcal{J}^i}$ has full rank. Therefore, \eqref{theorem_map} is obtained from solving \eqref{pa}. Substituting \eqref{theorem_map} into \eqref{pb}, we have \eqref{theorem_region}. 
\end{proof}
For the the linear programs \eqref{4c}-\eqref{4e} and \eqref{4f}-\eqref{4h} at the lower level,
Theorem \ref{Theorem 1} shows that given the domain $\Lambda$ of the parameters $\hat{y}, l, y$, the relationship between the optimal primal solutions $\bm{x}_{D}^{*}(\hat{y},l)$ and $\bm{z}_{R}^{*}(\hat{y},y)$, and the parameters $\hat{y}, l, y$ is \emph{piecewise linear} across $\Lambda$.  Concretely, the domain $\Lambda$ is defined as $\{0 \leq \hat{y} \leq C,l_{min} \leq l \leq l_{max},0 \leq y \leq C\}$, where $C$ is the capacity of wind power, and $l_{min},l_{max}$ are the minimum and maximum values of the load. Also, $\Lambda$ can be divided into a set of critical regions, where such relationship is affine. An illustration is shown in Fig. \ref{illustrationofmpp}. In this context, the derivation of loss function is based on the collection of such affine relationship, defined within the respective region. 

% This transformation depends on the KKT conditions of the lower level problem, which may exhibit varying formulations, reflecting the binding or non-binding nature of different constraints, under different parameter values, i.e., $\hat{y}, l, y$. Therefore, it is necessary to identify each region defined by $\hat{y}, l, y$ in which the KKT conditions remain unchanged.  In this context, the loss function comprises a collection of loss functions defined within their respective regions.

The procedure of obtaining the loss function is summarized into three steps. In the first step, we derive the relationship between the parameters $\hat{y},l,y$ and the lower level optimal solutions $\bm{x}_{D}^{*}(\hat{y},l)$ and $\bm{z}_{R}^{*}(\hat{y},y)$ in a critical region. By incorporating this relationship into the upper-level objective, we derive the loss function for the forecast model training that applies within the specific region in the second step. Finally, by assembling all the regions and their corresponding loss functions, we obtain the complete loss function across the entire domain.

% The bilevel program in \eqref{4} is hard to solve. \cite{zhang2023valueoriented} proposed a solution strategy by iteratively solving the upper level and the lower level problems. Under such an iterative scheme, the upper level objective is used as the loss function. However, such a loss function is only defined in the region around each training sample. Here, we propose to 
% derive the loss function across the domain, i.e., from zero to the wind power capacity, by collecting all the loss functions defined in the corresponding regions. To achieve this, we transform the bilevel program in \eqref{4} into a single-level program, by explicitly deriving the relationship between the forecasts and the lower level optimal solutions. We firstly derive the local relationship in a three-dimensional region defined by $\hat{y},l,y$. By plugging such local relationship into the upper level objective, the loss functions is obtained. Then, we show how to find all regions and corresponding loss functions, whose collection is exactly the loss function across the domain. An illustration of those steps is shown in Fig. \ref{steps}.

\subsection{Step 1: Deriving the Relationship between Parameters and Optimal Solutions in One Critical Region}
The first step is to establish the relationship between the parameters $\hat{y},l,y$ and the lower level optimal solutions $\bm{x}_D^*(\hat{y},l), \bm{z}_R^{*}(\hat{y},y)$ in a critical region. Here, we rewrite the day-ahead \eqref{4c}-\eqref{4e} and real-time \eqref{4f}-\eqref{4h} operation problems in the same form of \eqref{lp}. %The sample index $m$ is dropped for notational simplicity.  
\begin{subequations}\label{6}
\begin{align}
\bm{x}_{D}^*(\hat{y},l)=& \,\mathop{\arg\min}_{\bm{x}_{D}}  \bm{c}_D^\top  \bm{x}_{D}\label{6a}\\
     \text{s.t.} \quad  &  \bm{G}_D  \bm{x}_{D}\leq \bm{w}_D+\bm{F}_D\begin{bmatrix}
l-\hat{y}\\
-l+\hat{y}
\end{bmatrix}\label{6b}
\end{align}
\end{subequations}
and
\begin{subequations}\label{7}
\begin{align}
\bm{z}_{R}^*(\hat{y},y)=& \,\mathop{\arg\min}_{\bm{z}_{R}}  \bm{c}_R^\top  \bm{z}_{R}\label{7a}\\
     \text{s.t.} \quad  &  
    \bm{G}_{R}  \bm{z}_{R}\leq \bm{w}_{R}+\bm{F}_R\begin{bmatrix}
\hat{y}-y\\
y-\hat{y}
\end{bmatrix},\label{7b}
\end{align}
\end{subequations}

where 
\begin{equation*}
\begin{split}
& \bm{G}_D=\begin{bmatrix}
\bm{A}_{D}\\
\bm{1}^\top\\
-\bm{1}^\top
\end{bmatrix}
\bm{w}_D=\begin{bmatrix}
\bm{b}_{D}\\
0\\
0
\end{bmatrix}
\bm{F}_D=\begin{bmatrix}
\bm{O}\\
\bm{I}_{2\times2}
\end{bmatrix}\\
& \bm{G}_R=\begin{bmatrix}
\bm{A}_{R}\\
\bm{d}^{\top}_R\\
-\bm{d}^{\top}_R
\end{bmatrix}
\bm{w}_R=\begin{bmatrix}
\bm{b}_{R}\\
0\\
0
\end{bmatrix}
\bm{F}_R=\begin{bmatrix}
\bm{O}\\
\bm{I}_{2\times2}
\end{bmatrix}
\end{split}
\end{equation*}

Here, $\bm{O}$ is all-zero matrix, and $\bm{I}_{2\times2}$ is identity matrix.

% \begin{subequations}\label{6}
% \begin{align}
% \bm{x}_{D}^*(\hat{y},l),\bm{z}_{R}^*(\hat{y},y)=& \,\mathop{\arg\min}_{\bm{x}_{D},\bm{z}_{R}}  \bm{c}_D^\top  \bm{x}_{D}+\bm{c}_R^\top  \bm{z}_{R}\label{6a}\\
%      \text{s.t.} \quad  &  \bm{A}_D  \bm{x}_{D}\leq \bm{b}_D: \bm{\delta}\label{6b}\\
%      \quad & \bm{1}^\top  \bm{x}_{D}+\hat{y}=l:\lambda\label{6c}\\
%     \quad&
%     \bm{A}_{R}  \bm{z}_{R}\leq \bm{b}_{R}:\bm{\mu}\label{6d}\\
%      \quad & \bm{d}_{R}^\top  \bm{z}_{R}+y_{}-\hat{y}=0:\nu,\label{6e}
% \end{align}
% \end{subequations}
% where the dual variables are given after the colons. 

% \todo{what does the superscript $i$ mean in $\hat{y}^i,l^i, \mathcal{J}_D^i, \mathcal{J}_R^i$ etc? What is the difference of $\mathcal{J}_D^i$ and $\overline{\mathcal{J}_D}^i$?}
% \todo{Also, I didn't quite follow Proposition 1-3. It might be helpful to add more explanations after each proposition to interpret the results}

With \eqref{theorem_map} and \eqref{theorem_region} in Theorem \ref{Theorem 1}, we give the relationship in each critical region, for the day-ahead problem in Proposition~\ref{prop1} and for the real-time problem in Proposition~\ref{prop2}.
\begin{proposition}\label{prop1}
Let $\mathcal{J}_D$ denote the set of constraint indices in \eqref{6b}. Given the parameters $\hat{y},l \in \mathcal{B}_D^i$, let $(\mathcal{J}_D^i,\overline{\mathcal{J}_D}^i)$ be the optimal partition of $\mathcal{J}_D$ associated with $\hat{y},l \in \mathcal{B}_D^i$. Let $\bm{G}_{D,\mathcal{J}_D^i}$,$\bm{F}_{D,\mathcal{J}_D^i}$, and $\bm{w}_{D,\mathcal{J}_D^i}$ be respectively the submatrix of $\bm{G}_D$,$\bm{F}_D$ and the subvector of $\bm{w}_D$  corresponding to the index set $\mathcal{J}_D^i$. Let $\bm{G}_{D,\overline{\mathcal{J}_D}^i}$, $\bm{F}_{D,\overline{\mathcal{J}_D}^i}$ $\bm{w}_{D,\overline{\mathcal{J}_D}^i}$ be respectively the submatrix of $\bm{G}_D$,$\bm{F}_D$ and the subvector of $\bm{w}_D$  corresponding to the index set $\overline{\mathcal{J}_D}^i$. Assume that \eqref{6} is neither primal nor dual degenerate. The relationship is derived as,
\begin{equation}\label{map_da}
    \bm{x}_D^i(\hat{y},l) = \bm{G}_{D,\mathcal{J}_D^i}\\^{-1}(\bm{w}_{D,\mathcal{J}_D^i}+\bm{F}_{D,\mathcal{J}_D^i}\begin{bmatrix}
l-\hat{y}\\
-l+\hat{y}
\end{bmatrix}),
\end{equation}
where the associated critical region $\mathcal{B}_D^i$ is given by,
\begin{equation}\label{region_da}
\begin{split}
    &\bm{G}_{D,\overline{\mathcal{J}_D}^i}\bm{G}_{D,\mathcal{J}_D^i}^{-1}(\bm{w}_{D,\mathcal{J}_D^i}+\bm{F}_{D,\mathcal{J}_D^i}\begin{bmatrix}
l-\hat{y}\\
-l+\hat{y}
\end{bmatrix})\\
&< \bm{w}_{D,\overline{\mathcal{J}_D}^i}+\bm{F}_{D,\overline{\mathcal{J}_D}^i}\begin{bmatrix}
l-\hat{y}\\
-l+\hat{y}
\end{bmatrix}.
    \end{split}
    \end{equation}

\end{proposition}

\begin{proposition}\label{prop2}
Let $\mathcal{J}_R$ denote the set of  constraint indices in \eqref{7b}. Given the parameters $\hat{y},y \in \mathcal{B}_R^k$, let $(\mathcal{J}_R^k,\overline{\mathcal{J}_R}^k)$ be the optimal partition of  $\mathcal{J}_R$ associated with $\hat{y},y$. Let $\bm{G}_{R,\mathcal{J}_R^k}$, $\bm{F}_{R,\mathcal{J}_R^k}$, and $\bm{w}_{R,\mathcal{J}_R^k}$ be respectively the submatrix of $\bm{G}_R$,$\bm{F}_R$ and the subvector of $\bm{w}_R$  corresponding to the index set $\mathcal{J}_R^k$. Let $\bm{G}_{R,\overline{\mathcal{J}_R}^k}$, $\bm{F}_{R,\overline{\mathcal{J}_R}^k}$, $\bm{w}_{R,\overline{\mathcal{J}_R}^k}$ be respectively the submatrix of $\bm{G}_R$,$\bm{F}_R$ and the subvector of $\bm{w}_R$  corresponding to the index set $\overline{\mathcal{J}_R}^k$. Assume that \eqref{7} is neither primal nor dual degenerate. The relationship is derived as,
\begin{equation}\label{map_re}
    \bm{z}_R^k(\hat{y},y) = \bm{G}_{R,\mathcal{J}_R^k}\\^{-1}(\bm{w}_{R,\mathcal{J}_R^k}+\bm{F}_{R,\mathcal{J}_R^k}\begin{bmatrix}
\hat{y}-y\\
-\hat{y}+y
\end{bmatrix}),
\end{equation}
where the associated critical region $\mathcal{B}_R^k$ is given by,
\begin{equation}\label{region_re}
\begin{split}
&\bm{G}_{R,\overline{\mathcal{J}_R}^k}\bm{G}_{R,\mathcal{J}_R^k}^{-1}(\bm{w}_{R,\mathcal{J}_R^k}+\bm{F}_{R,\mathcal{J}_R^k}\begin{bmatrix}
\hat{y}-y\\
-\hat{y}+y
\end{bmatrix})\\
&< \bm{w}_{R,\overline{\mathcal{J}_R}^k}+\bm{F}_{R,\overline{\mathcal{J}_R}^k}\begin{bmatrix}
\hat{y}-y\\
-\hat{y}+y
\end{bmatrix}.
    \end{split}
    \end{equation}
\end{proposition}

The proof of two propositions is in Appendix \ref{Appendix A}. 

Let $\{\mathcal{B}_D^i\}_{i=1}^{K_D},\{\mathcal{B}_R^k\}_{k=1}^{K_R}$ be the set of critical regions in day-ahead and real-time problems, where $K_D,K_R$ denote the number of critical regions. In Proposition \ref{prop3}, we give the set of the critical regions of jointly considering those two problems.

\begin{proposition}\label{prop3}
Consider the set of the critical regions of jointly considering the day-ahead and real-time problems, i.e., $\{\mathcal{B}^{ik}\},\forall i=1,...,K_D,\forall k=1,...,K_R$. It is obtained by the Cartesian product of $\{\mathcal{B}_D^i\}_{i=1}^{K_D},\{\mathcal{B}_R^k\}_{k=1}^{K_R}$, i.e.,
\begin{equation}\label{region}
\begin{split}  
  \{\mathcal{B}^{ik}\}= & \{\mathcal{B}_D^i\}_{i=1}^{K_D} \times \{\mathcal{B}_R^k\}_{k=1}^{K_R}\\
  =& \{(\mathcal{B}_D^i,\mathcal{B}_R^k):\mathcal{B}_D^i\in \{\mathcal{B}_D^i\}_{i=1}^{K_D}, \mathcal{B}_R^k\in \{\mathcal{B}_R^k\}_{k=1}^{K_R}\}
\end{split}
\end{equation}
where the size of the set is $K_D\times K_R$

% The critical region $\mathcal{B}^ij$, that contains $\hat{y}^i,l^i,y^i$ and \eqref{map_da},  \eqref{map_re} hold, is,
% \begin{equation}\label{region}
% \begin{split}
%     &\bm{G}_{D,\overline{\mathcal{J}_D}^i}\bm{G}_{D,\mathcal{J}_D^i}^{-1}(\bm{w}_{D,\mathcal{J}_D^i}+\bm{F}_{D,\mathcal{J}_D^i}\begin{bmatrix}
% l-\hat{y}\\
% -l+\hat{y}
% \end{bmatrix})\\
% &< \bm{w}_{D,\overline{\mathcal{J}_D}^i}+\bm{F}_{D,\overline{\mathcal{J}_D}^i}\begin{bmatrix}
% l-\hat{y}\\
% -l+\hat{y}
% \end{bmatrix}\\
% &\bm{G}_{R,\overline{\mathcal{J}_R}^i}\bm{G}_{R,\mathcal{J}_R^i}^{-1}(\bm{w}_{R,\mathcal{J}_R^i}+\bm{F}_{R,\mathcal{J}_R^i}\begin{bmatrix}
% \hat{y}-y\\
% -\hat{y}+y
% \end{bmatrix})\\
% &< \bm{w}_{R,\overline{\mathcal{J}_R}^i}+\bm{F}_{R,\overline{\mathcal{J}_R}^i}\begin{bmatrix}
% \hat{y}-y\\
% -\hat{y}+y
% \end{bmatrix}.
%     \end{split}
%     \end{equation}
\end{proposition}

\subsection{Step 2: The Loss Function in One Region}
Based on Propositions \ref{prop1}-\ref{prop3}, we can use the mapping of $\bm{x}_D^i(\hat{y},l)$ and $\bm{z}_R^k(\hat{y},y)$ to reduce the bilevel program in \eqref{4} into a single-level one. Specifically, the objective function of the reduced single-level program is,
\begin{equation}\label{loss function}
\ell^{ik}(\hat{y},l,y;\mathcal{B}^{ik})=\bm{c}_D^\top \bm{x}_D^i(\hat{y},l)+\bm{c}_R^\top \bm{z}_R^k(\hat{y},y),
\end{equation}
which is an affine function defined in the region $\mathcal{B}^{ik}$ where \eqref{map_da} and \eqref{map_re} hold. For clarity, we rewrite $\ell^{ik}(\hat{y},l,y;\mathcal{B}^{ik})$ in the explicit form of $\hat{y},l,y$,
\begin{equation}\label{loss rewrite}
    \begin{split}
        \ell^{ik}(\hat{y},l,y;\mathcal{B}^{ik})=\beta_{\hat{y}}^{ik}\hat{y}+\beta_{l}^{ik}l+\beta_{y}^{ik}y+\beta_0^{ik},
    \end{split}
\end{equation}
where $\beta_{\hat{y}}^{ik}$,$\beta_{l}^{ik}$,$\beta_{y}^{ik}$, and $\beta_0^{ik}$ are the coefficients obtained from \eqref{map_da} and \eqref{map_re}. Eq. \ref{loss rewrite} maps the parameters to the operation cost, and defines the loss function in a critical region.

\begin{figure}[h]
  \centering
  \includegraphics[scale=0.6]{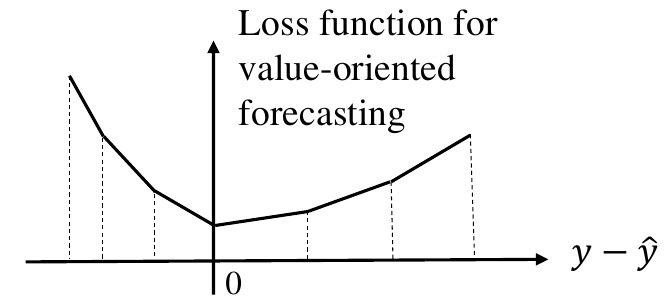}\\
  \caption{An illustration of loss function w.r.t. the deviation $y-\hat{y}$, with fixed values of $l,y$.}\label{Fig 2}
\end{figure}

\subsection{Step 3: The Derived Loss Function}
Then we move to find all critical regions and the corresponding loss functions. The main steps are summarized in Algorithm \ref{alg1}. We start with one critical region $\mathcal{B}^{ik}$, and find the corresponding loss function from \eqref{loss rewrite}. We then keep exploring the remaining region $\Lambda-\mathcal{B}^{ik}$, with the similar procedures of identifying the relationship in \eqref{loss rewrite} and region defined by \eqref{region}. %Then, the loss function $\ell^{ik}(\hat{y},l,y;\mathcal{B}^{ik})$ in \eqref{loss rewrite} is formulated. 
The algorithm terminates when the entire parameter space $\Lambda$ is covered. 
\begin{algorithm}[h]
	%\textsl{}\setstretch{1.8}
	%\renewcommand{\algorithmicrequire}{\textbf{Input:}}
	\caption{The derivation of loss function for value-oriented forecasting}
	\label{alg1}
	\begin{algorithmic}[1]
    \Require{The initial region of parameters $\Lambda$}
    \State{In the given set of $\Lambda$, solve \eqref{6},\eqref{7} by treating $\hat{y},l,y$ as free variables to obtain feasible points.}
    \State{Obtain $\bm{x}_D^i(\hat{y},l)$ by \eqref{map_da} and $\bm{z}_R^k(\hat{y},y)$ by \eqref{map_re}, and the region $\mathcal{B}^{ik}$ by \eqref{region}.}
    \State{Obtain the loss function $\ell^{ik}(\hat{y},l,y;\mathcal{B}^{ik})$ defined in \eqref{loss rewrite}}. 
    \State{Obtain the remaining region $\Lambda-\mathcal{B}^{ik}$.}
    \State{If no more regions to explore, go to next step; otherwise given the remaining region, solve \eqref{6},\eqref{7} by treating $\hat{y},l,y$ as free variables to obtain new feasible points, and then go to Step 2.}
    \State{Collect all the loss functions and unify the critical regions.}
	\end{algorithmic}  
\end{algorithm}

To sum up, the domain $\Lambda$ can be divided into several critical regions $\{\mathcal{B}^{ik}\}$, where the loss function $\ell^{ik}(\hat{y},l,y;\mathcal{B}^{ik})$ is affine. The collection of the loss functions $\{\ell^{ik}(\hat{y},l,y;\mathcal{B}^{ik})\}$ gives a piecewise linear function. An illustration of such a function with respect to the deviation $y-\hat{y}$, with fixed values of $l,y$, is shown in Fig. \ref{Fig 2}. Ultimately, we can estimate the parameters $\Theta$ of the forecast model $g$, i.e.,
\begin{subequations}\label{8}
\begin{alignat}{2}
     &\hat{\Theta}=&& \quad\mathop{\arg\min}_{\Theta}
      \frac{1}{M}\sum_{m=1}^M [\ell^{ik}_{(m)}(\hat{y}_m,l_m,y_m;\mathcal{B}^{ik}_{(m)})]\label{8a}\\
    &\text{s.t.} && \quad \hat{y}_m=g(\bm{s}_m;\Theta),\label{8b}\\
    &&& \quad
    \hat{y}_m,l_m,y_m \in \mathcal{B}^{ik}_{(m)}.
\end{alignat}
\end{subequations}
where we use $\mathcal{B}^{ik}_{(m)}$ to denote the critical region that the sample indexed by $m$ belong to, and the corresponding loss function is denoted by $\ell^{ik}_{(m)}(\hat{y}_m,l_m,y_m;\mathcal{B}^{ik}_{(m)})$.
We can train different types of forecast model via \eqref{8}. Here we present a neural network (NN) forecast model with batch optimization as an example. The training stage is presented in Algorithm \ref{alg2}.
\begin{algorithm}
	%\textsl{}\setstretch{1.8}
	%\renewcommand{\algorithmicrequire}{\textbf{Input:}}
	\caption{Training stage of a neural network for value-oriented forecasting}
	\label{alg2}
	\begin{algorithmic}[1]
	\Require{Learning rate $\alpha$, batch size $Q$, and initialized forecast model parameters $\hat{\Theta}^0$}
	\For{epoch $e =1,2, \ldots$}
	
        \State{Sample batch from the training set: $\{\bm{s}_{m},y_{m}\}_{m=1}^Q$}
	    \State{Output the forecasts via model $g$ parameterized by $\hat{\Theta}^{e-1}$: $\{\hat{y}_{m}=g(\bm{s}_{m};\hat{\Theta}^{e-1})\}_{m=1}^Q$}
	  
	    \For{data $m = 1, \ldots,Q$}
	        \State{Find the critical region $\mathcal{B}^{ik}_{(m)}$ that $\hat{y}_{m},l_m,y_m$ belong to and calculate the corresponding loss function $\ell^{ik}_{(m)}(\hat{y}_m,l_m,y_m;\mathcal{B}^{ik}_{(m)})$ defined in \eqref{loss rewrite}
        }
	   \EndFor
	   \State{Update forecast model parameters via gradient descent: 
    $$\hat{\Theta}^e \leftarrow \hat{\Theta}^{e-1}-\alpha \bigtriangledown_{\hat{\Theta}^{e-1}}\sum_{m=1}^Q\ell^{ik}_{(m)}(\hat{y}_m,l_m,y_m;\mathcal{B}^{ik}_{(m)})$$}
    \EndFor
\end{algorithmic} \end{algorithm}

% Line 3 - Line 6: Obtain the batch of forecasts with the estimated model parameters $\hat{\Theta}^{e-1}$ in the last epoch. Find the region that the sample $\hat{y}_m,l_m,y_m,\forall m=1,..,Q$ belong to and calculate the loss function for it.

\subsection{Relationship to the Solution Perspective based on Differential Optimization}
Here, we discuss the relationship between the proposed approach and the differential optimization approach in \cite{donti2017task}, where no loss function is explicitly derived. Concretely, \cite{donti2017task} proposes to encode operation problem as a differentiable optimization layer, and a forecast model is trained together with the differentiable optimization layer in an end-to-end fashion. For each model training iteration, it consists a forward and a backward process. In the forward process, it solves the day-ahead and real-time energy dispatch problems and obtain the optimal solutions and operation costs, for each data sample. In the backward process, it calculates the derivative of the optimal lower level solutions w.r.t. the forecast, i.e., $\frac{\partial \bm{x}_D^*(\hat{y},l)}{\partial \hat{y}}$ and $\frac{\partial \bm{z}_R^*(\hat{y},y)}{\partial \hat{y}}$ based on the KKT conditions of \eqref{6} and \eqref{7} using automatic differentiation. The gradient $\bm{c}_D^\top\frac{\partial \bm{x}_D^*(\hat{y},l)}{\partial \hat{y}}+\bm{c}_R^\top\frac{\partial \bm{z}_R^*(\hat{y},y)}{\partial \hat{y}}$ is then used for updating the NN-based forecast model. Based on \eqref{loss rewrite}, the gradient $\bm{c}_D^\top\frac{\partial \bm{x}_D^*(\hat{y},l)}{\partial \hat{y}}+\bm{c}_R^\top\frac{\partial \bm{z}_R^*(\hat{y},y)}{\partial \hat{y}}$ is the slope of $\hat{y}$, i.e., $\beta_{\hat{y}}^{ik}$.
% When we use the loss function in \eqref{newloss}, the gradient of $\frac{\partial \ell^i(\hat{y},y;\mathcal{B}^i)}{\partial \hat{y}}$ is also $\beta_{\hat{y}}^i$. 
In this regard, the two methods are equivalent, when training a NN-based forecast model. 

However, explicitly deriving the loss function for value-oriented forecasting can bring benefits from two sides, compared to the end-to-end training approach with differentiable optimization. Firstly, it enables the utilization of regression models beyond neural networks, expanding the applicability of different types of forecast models such as the models based on decision tree like XGBoost and LightGBM. Secondly, the presence of the derived loss function enhances computational efficiency. In contrast, the approach presented in \cite{donti2017task} needs to solve the optimization problem in the forward path and computes the derivation of KKT conditions for \emph{each} sample during \emph{every} training epoch, which poses a greater computational challenge.

\section{Case Study}
We consider the operation of a VPP operator with two SGs, flexible resources, and a wind power generator. The power generation plan of SGs is determined through solving the day-ahead problem denoted as \eqref{2},  taking into account the wind power forecast $\hat{y}$. After the wind power realization $y$ is revealed, the operator solves \eqref{3} and utilizes flexible resources to address energy imbalances. The detailed parameter settings are the same as \cite{zhang2023valueoriented}. Multi-layer perceptron (MLP) is used as the forecast model, whose hyper-parameters are summarized in Table \ref{mlp}. Its input is the contextual information formed by the estimated wind speed and direction at 10m and 100m altitude, and the output is the forecast in the corresponding hour. Wind and load data are available in \cite{ppm2022}\footnote{Code will be available after publication.}. In this study, we employ a quality-oriented forecasting approach, whose training stage is shown in \eqref{mse}, as the benchmark for comparison.
\begin{table}[h]
\caption{Summary of MLP forecast model hyper-parameters}\label{mlp}
\begin{center}
\begin{tabular}{ c  c }
\hline\hline
    Description & Value\\
\hline
    No. of hidden layers & 2\\
    No. of neurons in hidden layer & 256\\
    Activation function in hidden layer & Relu\\
    Activation function in output layer & Sigmoid\\
    Dim. of contextual information & 4 \\
    % Output dim. & 1\\
    Batch size ($Q$) and Optimizer & 512, Adam\\
    Learning rate $\alpha$ & 1e-3\\
\hline\hline
\end{tabular}
\end{center}
\end{table}

To evaluate the forecast quality and value on test set, we use root mean squared error (RMSE) and average monetary score (AMS) as evaluation metrics. The RMSE is defined by 

\begin{equation}
    \text{RMSE} = \sqrt{\frac{1}{M_{test}}\sum_{m=1}^{M_{test}}(y_m-\hat{y}_m)^2},
\end{equation}
where $M_{test}$ is the number of the samples in the test set.
We define the average monetary score to measure the value of forecast to the operation problem, which is defined as
\begin{equation}\label{money}
\text{AMS} = \frac{1}{M_{test}}\sum_{m=1}^{M_{test}}\bm{c}_D^\top \bm{x}_D^*(\hat{y}_m,l_m)+\bm{c}_R^\top \bm{z}_R^*(\hat{y}_m,y_m).
\end{equation}
The lower the average monetary score, the better the value.

The performance analysis is approached from three key aspects. First, in order to enhance the interpretation of the loss function, we visualize it.
Second, to demonstrate its operational superiority, we conduct a comparative analysis between the proposed approach and quality-oriented forecasting. This analysis focuses on the average monetary score, as defined in equation \eqref{money}, across varying levels of wind penetration. Lastly, to showcase its computational efficiency, we compare it with the differential optimization detailed in reference \cite{donti2017task}, which lacks an explicit definition of the loss function.

\subsection{The Visualization of the Loss Function}
\begin{figure}[t]
  \centering
  \includegraphics[scale=0.6]{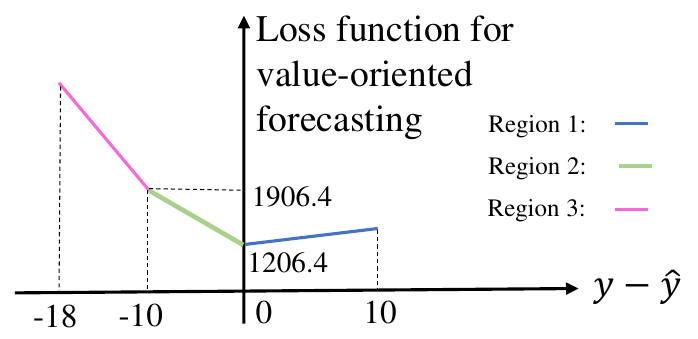}\\
  \caption{The loss function within three regions.}\label{Fig 3}
\end{figure}

In this subsection, the wind capacity is set as \SI{28}{\kW}, which means $0 \leq y,\hat{y} \leq 28$. The loss function is a piecewise linear function defined within three regions. The loss funciton defined in \eqref{loss rewrite} is written as,
\begin{equation}
\ell^{ik}(\hat{y},l,y;\mathcal{B}^{ik})=
    \begin{cases}
        -10\hat{y}+30l-20y+6.4 & \text{Region 1}\\
        70\hat{y}+30l-100y+6.4 & \text{Region 2}\\
        170\hat{y}+30l-200y-993.6 & \text{Region 3}.
    \end{cases}
\end{equation}
We fix the values of $y,l$ as \SI{10}{\kW},\SI{50}{\kW}, and plot such loss functions w.r.t. the deviation $y-\hat{y}$ in Fig. \ref{Fig 3}, where the loss functions in different regions are plotted with different color. It is convex with two segments on the left plane and one segment on the right plane. The loss achieves the minimum when there is no deviation, i.e., $\hat{y}=y$. When the deviation $y-\hat{y}$ is larger than \SI{-10}{\kW} but smaller than 0 (the region 2), i.e. $\hat{y}-10 \leq y \leq \hat{y}$, the VPP operator can use the flexible resources with cheaper marginal cost to settle the deviation. When the deviation $y-\hat{y}$ is smaller than \SI{-10}{\kW}, the flexible resources with larger marginal cost is used. Therefore, the green segment is more flatten than the pink one. %\todo{discuss about Fig. 4}

\subsection{The Operational Advantage under Different Levels of Wind Power Penetration}
Here we compare the performance of the same MLP forecast models whose training stage is in \eqref{mse} and \eqref{4} respectively. We denote the forecast issued by the former model as \textit{Quality-oriented forecast} and that by the latter model as \textit{Value-oriented forecast} in Table \ref{Table 2} and \ref{Table 3}. We test their performance under three different wind power capacity levels,
10, 20, 28 \unit{\kW}. %Under different  the results of value-oriented and quality-oriented forecasts are reported in Table \ref{Table 2} and \ref{Table 3}. 
Table \ref{Table 2} shows the RMSE under different wind power capacities, and Table \ref{Table 3} shows the average monetary score. We can see that although value-oriented forecasting has higher higher forecasting error compared to the quality-oriented one, it achieves lower operation cost. The results highlight the fact that the good statistical quality of forecast does not necessarily ensure the good value in the operation. Therefore, we need to design value-oriented forecasting approach that captures the value of forecast in the actual operation problems. Also, with the increase penetration of wind power, the cost reduction, resulted by value-oriented forecasting, is more obvious, which indicates the value-oriented forecasting is more preferred under large penetration of wind.
\begin{table}[h]
\caption{RMSE under different wind power capacity}\label{Table 2}
\begin{center}
\begin{tabular}{ c  c  c  }
\hline\hline
         \makecell[c]{Wind power \\capacity (\unit{\kW})} & \makecell[c]{Value-oriented\\ forecast (\unit{\kW})} & \makecell[c]{Quality-oriented\\ forecast (\unit{\kW})}\\
\hline
    \makecell[c]{10} & 2.6 & \textbf{1.8}\\
    \makecell[c]{20} & 5.3 & \textbf{3.7}\\
    \makecell[c]{28} & 7.5 & \textbf{5.1}\\
\hline\hline
\end{tabular}
\end{center}
\end{table}

\begin{figure}[h]
  \centering
  \includegraphics[scale=0.6]{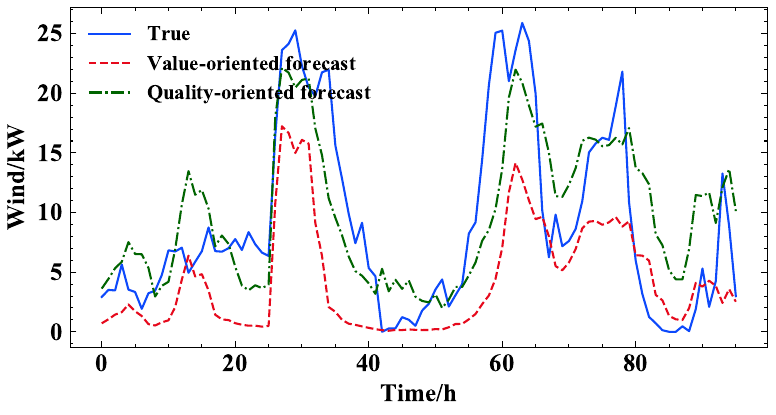}\\
  \caption{4-day wind power forecast profiles of the value-oriented and quality-oriented forecasting approaches.}\label{Fig 4}
\end{figure}

\begin{table}[h]
\caption{Average monetary score (AMS) under different wind power capacity}\label{Table 3}
\begin{center}
\begin{tabular}{ c  c  c  c }
\hline\hline
         \makecell[c]{Wind power\\
         capacity (\unit{kW})} & \makecell[c]{Value-oriented\\ forecast (\$) } & \makecell[c]{Quality-oriented\\ forecast (\$) } & \makecell[c]{Cost \\reduction (\$)}\\
\hline
    \makecell[c]{10} & \textbf{1569} & 1591 & -22\\
    \makecell[c]{20} & \textbf{1513} & 1558 & -45\\
    \makecell[c]{28} & \textbf{1466} & 1535 & -69\\
\hline\hline
\end{tabular}
\end{center}
\end{table}
Under the wind power capacity of \SI{28}{\kW}, we show the 4-day forecast results of the quality- and value-oriented forecasting approaches in Fig \ref{Fig 4}. The loss function has a clear impact on the forecast model performance. The forecast model tends to predict less power than the quality-oriented one, since in in this case, the cost for energy deficit (where the forecast is larger than the real-time realization) is more expensive than the cost for energy surplus (where the forecast is smaller than the realization). %Therefore, the less profitable situation of energy deficit can happen less.

\subsection{Computational Complexity Comparison}

In this section, we compare the training time of the proposed value-oriented approach, the quality-oriented approach, and the differential optimization approach proposed in \cite{donti2017task} for value-oriented forecasting, all using the same MLP model (in Table \ref{mlp}) as forecast models.
%, which relies on the derivatives of the optimal solutions $\bm{x}_D^*(\hat{y},l),\bm{z}_R^*(\hat{y},y)$ with respect to the forecast $\hat{y}$ for updating the forecast model, is used as the baseline of the value-oriented forecasting. 
The training time is given in Table \ref{Table 4}. At the training stage, the proposed approach needs to find the region and the corresponding loss function. Therefore, it has longer training time than the quality-oriented one. However, its training time is much shorter than that of the differentiable optimization approach, which needs to repeatedly solve the energy dispatch problems and computes the derivatives for each sample. Therefore, by explicitly designing a loss function, the computational cost can be much lower.
\begin{table}[h]
\caption{Training time comparison}\label{Table 4}
\begin{center}
\begin{tabular}{ c  c c}
\hline\hline
     \makecell[c]{The proposed \\approach} & 
    \makecell[c]{Quality-oriented \\forecast} & 
    \makecell[c]{Baseline: differential \\optimization approach \cite{donti2017task}}\\
\hline
    1 min 4s  &3.18 s &23 min\\
\hline\hline
\end{tabular}
\end{center}
\end{table}

% Furthermore, we compare the average monetary score of the proposed  and the baseline approaches, under the wind power capacity of 28 $kW$. The average monetary score of the baseline approach is 2038 $\$$, which is larger than that of the proposed approach listed in Table \ref{Table 2}, i.e., 1466 $\$$. Without the global loss function, the baseline approach may easily get stuck in several regions, and cannot explore the whole picture. Therefore, the forecast model cannot effectively learn how to lower the operation cost.

\section{Conclusion}

In this paper, we theoretically derive the loss function for value-oriented wind power forecasting. At the training stage, a bilevel program is formulated, where the lower level solves the day-ahead and real-time operation problems, given the forecast model outputs provided by the upper level. The relationship between the lower level optimal solutions and the forecasts is theoretically derived. By substituting such relationship into the upper level objective for minimizing the expected operation cost, the loss function for value-oriented forecasting is obtained.

We evaluate the performance of the proposed approach via the day-ahead and real-time energy dispatch of a VPP operator in charge of wind power. The operations are in the form of linear program, and the derived value-oriented loss is a piecewise linear, nonnegative and convex function. Numerical studies show that although the issued forecasts by the proposed approach have larger RMSE, the average operation cost is smaller than that of the quality-oriented one. Also, by comparing to the value-oriented forecasting approach without explicitly deriving the loss function, the proposed approach is more computationally efficient. We note that the operation problems in different forms may have different loss functions. Here, the operation problem in the form of linear program is considered. In the future, it is interesting to derive loss function for operation problems in other forms.

\bibliographystyle{IEEEtran}
\bibliography{mylib}

\appendices
\section{Proof of Propositions 1, 2}\label{Appendix A}
\subsection{Proof of Proposition 1}
\begin{proof}

For the problem \eqref{6}, given the optimal partition $(\mathcal{J}_D^i,\overline{\mathcal{J}_D}^i)$, we have,
\begin{subequations}
\begin{align}
&\bm{G}_{D,\mathcal{J}_D^i}\bm{x}_D^i(\hat{y},l)=\bm{w}_{D,\mathcal{J}_D^i}+\bm{F}_{D,\mathcal{J}_D^i}\begin{bmatrix}
l-\hat{y}\\
-l+\hat{y}
\end{bmatrix}\label{partition_da}\\
& \bm{G}_{D,\overline{\mathcal{J}_D}^i}\bm{x}_D^i(\hat{y},l) < \bm{w}_{D,\overline{\mathcal{J}_D}^i}+\bm{F}_{D,\overline{\mathcal{J}_D}^i}\begin{bmatrix}
l-\hat{y}\\
-l+\hat{y}
\end{bmatrix} \label{partition_db}     
\end{align}
\end{subequations}
With Assumption \ref{assumption 2}, \eqref{6} is either primal nor dual degenerate, $\bm{G}_{D,\mathcal{J}_D^i}$ has full rank. Therefore, \eqref{map_da} is proved by \eqref{partition_da}.
By substituting \eqref{map_da} into \eqref{partition_db}, \eqref{region_da} is obtained.
\end{proof}

\subsection{Proof of Proposition 2}
\begin{proof}

For the problem \eqref{7}, given the optimal partition $(\mathcal{J}_R^k,\overline{\mathcal{J}_R}^k)$, we have,
\begin{subequations}
\begin{align}
&\bm{G}_{R,\mathcal{J}_R^k}\bm{z}_R^k(\hat{y},y)=\bm{w}_{R,\mathcal{J}_R^k}+\bm{F}_{R,\mathcal{J}_R^k}\begin{bmatrix}
\hat{y}-y\\
-\hat{y}+y
\end{bmatrix}\label{partition_ra}\\
& \bm{G}_{R,\overline{\mathcal{J}_R}^k}\bm{z}_R^k(\hat{y},y) < \bm{w}_{R,\overline{\mathcal{J}_R}^k}+\bm{F}_{R,\overline{\mathcal{J}_R}^k}\begin{bmatrix}
\hat{y}-y\\
-\hat{y}+y
\end{bmatrix} \label{partition_rb}     
\end{align}
\end{subequations}
With Assumption \ref{assumption 2}, \eqref{7} is either primal nor dual degenerate, $\bm{G}_{R,\mathcal{J}_R^i}$ has full rank. Therefore, \eqref{map_re} is proved by \eqref{partition_ra}.
By substituting \eqref{map_re} into \eqref{partition_rb}, \eqref{region_re} is obtained.
\end{proof}

\end{document}